\documentclass[a4paper,fleqn]{cas-dc}

\usepackage{amsmath,amssymb}
\usepackage{graphicx}
\usepackage{booktabs}
\usepackage{tabularx}
\usepackage{float}
\usepackage{capt-of}
\usepackage[section]{placeins}
\usepackage{siunitx}
\DeclareSIUnit{\clight}{\ensuremath{\mathit{c}}}
\usepackage{xcolor}
\usepackage{enumitem}
\usepackage{hyperref}
\hypersetup{hidelinks}
\usepackage[numbers,sort&compress]{natbib}
\ExplSyntaxOn
\cs_set:Npn \__first_footerline:
 {
  \group_begin:
  \small\sffamily
  \__short_authors:\; \rmfamily\itshape Preprint
  \group_end:
 }
\ExplSyntaxOff

\begin{document}
\renewcommand{\printorcid}{}
\let\WriteBookmarks\relax
\renewcommand{\topfraction}{0.95}
\renewcommand{\bottomfraction}{0.85}
\renewcommand{\textfraction}{0.05}
\renewcommand{\floatpagefraction}{0.80}
\renewcommand{\dbltopfraction}{0.95}
\renewcommand{\dblfloatpagefraction}{0.80}
\setcounter{topnumber}{3}
\setcounter{bottomnumber}{2}
\setcounter{totalnumber}{5}

\shorttitle{UVT acrylic Cherenkov counter for hadron identification}
\shortauthors{T. Gogami et al.}

\title[mode=title]{Beam performance and data-constrained optical modeling of a UV-transmitting acrylic Cherenkov counter for hadron identification}

\author[kyoto,shiga]{T. Gogami}
\cormark[1]
\ead{gogami.toshiyuki.u08@kyoto-u.jp}
\author[kyoto]{T. Taniguchi}
\author[tohoku]{D. Watanabe}
\author[korea]{J. K. Ahn}
\author[unito,infntorino]{E. Botta}
\author[jaeaasrc]{K. Ebata}
\author[infntorino]{A. Feliciello}
\author[scitokyo]{H. Fujioka}
\author[tokyo]{M. Fujita}
\author[kyoto,jaeaasrc]{T. K. Harada}
\author[tohoku]{S. H. Hayakawa}
\author[tohoku]{Y. Hong}
\author[tohoku]{Y. Ichikawa}
\author[tohoku]{R. Imamoto}
\author[kyoto]{K. Ishido}
\author[kyoto]{S. Iwai}
\author[kyoto]{T. Iwamoto}
\author[tohoku]{K. Miwa}
\author[kyoto]{T. Nagae}
\author[tohoku]{T. Nanamura}
\author[tohoku]{R. Sasaki}
\author[tohoku]{C. Seong}
\author[kyoto]{S. Shibata}
\author[kyoto]{S. Takahashi}
\author[tohoku]{H. Tamura}
\author[jaeaasrc]{K. Tanida}
\author[kyoto]{K. Tsutsumi}
\author[tohoku,kek]{M. Ukai}
\author[jaeaasrc]{T. O. Yamamoto}

\cortext[1]{Corresponding author}

\affiliation[kyoto]{organization={Graduate School of Science, Department of Physics, Kyoto University},
  city={Kyoto}, postcode={606-8502}, state={Kyoto}, country={Japan}}
\affiliation[shiga]{organization={Data Science and AI Innovation Research Promotion Center, Shiga University},
  city={Hikone}, postcode={522-8522}, state={Shiga}, country={Japan}}
\affiliation[tohoku]{organization={Department of Physics, Graduate School of Science, Tohoku University},
  city={Sendai}, postcode={980-8578}, country={Japan}}
\affiliation[korea]{organization={Department of Physics, Korea University},
  city={Seoul}, postcode={136-713}, country={Korea}}
\affiliation[unito]{organization={Department of Physics, University of Turin},
  addressline={Via P. Giuria 1}, city={Torino}, postcode={I-10125}, country={Italy}}
\affiliation[infntorino]{organization={Istituto Nazionale di Fisica Nucleare (INFN), Sezione di Torino},
  addressline={Via P. Giuria 1}, city={Torino}, postcode={I-10125}, country={Italy}}
\affiliation[jaeaasrc]{organization={Advanced Science Research Center, Japan Atomic Energy Agency},
  city={Tokai}, postcode={319-1195}, state={Ibaraki}, country={Japan}}
\affiliation[scitokyo]{organization={Department of Physics, School of Science, Institute of Science Tokyo},
  city={Tokyo}, postcode={152-8551}, country={Japan}}
\affiliation[tokyo]{organization={Department of Physics, Graduate School of Science, The University of Tokyo},
  city={Tokyo}, postcode={113-0033}, country={Japan}}
\affiliation[kek]{organization={Institute of Particle and Nuclear Studies, High Energy Accelerator Research Organization (KEK)},
  city={Tsukuba}, postcode={305-0801}, state={Ibaraki}, country={Japan}}

\begin{abstract}
A threshold Cherenkov counter based on ultraviolet-transmitting (UVT) acrylic has been developed for hadron identification in the S-2S spectrometer at J-PARC. Three \SI{730}{mm}$\times$\SI{100}{mm}$\times$\SI{50}{mm} radiators were beam tested: polished UVT acrylic without WLS, UVT acrylic roughened on the four long surfaces without WLS, and polished UVT acrylic containing 10 ppm by mass bis-MSB as a wavelength shifter (WLS). The polished/WLS prototype gave the largest two-ended photoelectron yield among the tested configurations. At a 1-photoelectron software-SUM threshold, its measured detection efficiency remained above \SI{99}{\percent} throughout the scanned longitudinal range, with no significant degradation up to an average extracted-particle rate of approximately $2.5\times10^{5}~\mathrm{s^{-1}}$ during beam delivery.

Beam observables constrained a Geant4 optical model with one common detector-response scale and no PMT-specific response tuning. The model reproduces the observed response hierarchy and main longitudinal trends. In the central $|y|<\SI{200}{mm}$ region, simulated and measured MPV estimators for the calibration configurations typically agree within about 15\%, while larger deviations occur toward the radiator ends, especially for rough/no-WLS. With the calibrated parameters fixed, the model was used to compare unmeasured detector configurations and quantify the momentum dependence of $K^+/p$ discrimination. In the central, normal-incidence geometry, a 90\% $K^+$-survival working point is predicted to keep proton survival below 3\% up to about \SI{1.8}{GeV/\clight}. The measurements and constrained extrapolations support the polished UVT/WLS configuration for S-2S and quantify its expected particle-identification reach beyond the simple threshold regime.
\end{abstract}


\maketitle

\section{Introduction}
\label{sec:intro}

High-resolution missing-mass spectroscopy with secondary hadron beams relies on efficient particle identification under high-rate conditions. A threshold Cherenkov counter can be used in two complementary ways: a positive signal can select a species above threshold while rejecting a heavier species below threshold, or the absence of a signal can retain a below-threshold species while vetoing lighter backgrounds. The latter logic is used, for example, by the aerogel Cherenkov counter in the S-2S particle-identification system to retain kaon candidates while suppressing pion and muon backgrounds \cite{Gogami2022S2S}. The useful operating range of a threshold Cherenkov counter is not restricted to this binary on/off regime. When both species radiate, differences in their Cherenkov yield can still be exploited by an appropriate discriminator or software threshold; conversely, at sufficiently high momentum, where both velocities approach unity and the light yields converge, the separation power must disappear. The momentum range over which a given radiator remains useful is therefore determined jointly by the Cherenkov thresholds, the detected photoelectron-count (NPE) distributions, and the chosen signal-efficiency operating point.

S-2S is a magnetic spectrometer at J-PARC composed of normal-conducting quadrupole and dipole magnets together with downstream tracking and particle-identification detectors \cite{Gogami2022S2S}. For acrylic with refractive index $n\simeq1.49$, the Cherenkov thresholds are approximately \SI{0.45}{GeV/\clight} for $K^+$ and \SI{0.85}{GeV/\clight} for protons. The J-PARC E94 $({\pi}^+,K^+)$ program provides a direct application of this detector concept. For a candidate S-2S setting with central momentum $p_{\rm cent}\simeq\SI{0.72}{GeV/\clight}$, the scattered-kaon acceptance is approximately \SIrange{0.61}{0.83}{GeV/\clight} for a \SI{1.05}{GeV/\clight} incident pion beam \cite{Gogami2022S2S,E94Proposal}. In this particular setting, kaons radiate while primary protons remain below threshold, making an acrylic counter naturally suited to kaon selection and proton rejection. More generally, the same detector concept can remain useful above the proton threshold through differences in detected light yield, as previously exploited in Cherenkov-based hadron identification \cite{Gogami2013Bucking,Gogami2016WC}. In a full spectrometer, this information can be combined with time-of-flight and tracking for trigger and offline particle identification \cite{Gogami2018HypernuclearTechniques}. The present work concentrates on the intrinsic response of one counter module and its optical-model description. System-level performance of the installed 16-module S-2S detector, including multiple particle species and experiment-specific conditions, will be characterized separately.

The detector challenge is not set by the refractive index alone. Cherenkov emission is strongly weighted toward short wavelength,
\begin{equation}
 \frac{\mathrm{d}^2N}{\mathrm{d}x\,\mathrm{d}\lambda}
 = \frac{2\pi\alpha Z_{\rm ch}^{2}}{\lambda^2}
 \left(1-\frac{1}{\beta^2n^2(\lambda)}\right),
\end{equation}
where $\alpha$ is the fine-structure constant, $Z_{\rm ch}$ is the particle charge in units of the elementary charge, $\beta=v/c$, $\lambda$ is the photon wavelength, and $n(\lambda)$ is the wavelength-dependent refractive index. Thus, the useful signal depends on the coupled effects of ultraviolet transmission, bulk absorption, optical boundaries, wavelength conversion, and two-ended collection over a long acrylic path. UV-transmitting (UVT) acrylics have been characterized for Cherenkov applications and show substantial material-dependent differences in the ultraviolet region \cite{Zwinkels1990UVTAcrylic}. Earlier S-2S detector development likewise demonstrated the practical importance of UVT acrylic components and UV-sensitive photomultiplier tubes (PMTs) \cite{Gogami2016WC}. In the present counter, the UVT acrylic is itself the radiator, making its bulk transmission directly relevant to the detected Cherenkov spectrum.

Wavelength shifting provides a second route to improved photon collection. Wavelength-shifting acrylic has previously been studied both for Cherenkov light produced inside the acrylic and for externally generated Cherenkov light entering the material \cite{Carl2004WLSAcrylic,Beckford2014WLSAcrylic}. Bis-MSB absorbs short-wavelength photons and re-emits them at longer wavelength, where acrylic transmission and photodetector sensitivity can be more favorable. This process also changes the photon direction and can therefore modify the longitudinal response independently of the bulk attenuation. Surface roughening offers a qualitatively different mechanism: it redistributes photon directions at optical boundaries without changing the photon wavelength, an approach also used to recover light collection in a recent low-$\beta$ fused-silica threshold-Cherenkov counter \cite{Saito2026KVC}. The combined effects of bulk UV transmission, surface treatment, and wavelength shifting have not, however, been quantified for the present long acrylic threshold-radiator geometry with two-ended readout.

To compare these effects experimentally, three UVT-acrylic prototypes with identical $730\times100\times50~\mathrm{mm^3}$ radiator geometry were constructed: polished without wavelength shifter, roughened on the four long surfaces without wavelength shifter, and polished with bis-MSB at a concentration of 10 parts per million (ppm) by mass. Their absolute photoelectron yield, longitudinal response, two-ended efficiency, and rate stability were measured with hadron beams. The \SI{730}{mm} transport length makes photon collection over several hundred millimeters an integral part of the detector response, rather than a property that can be inferred from small material samples alone.

This work combines beam measurements of three UVT-acrylic prototypes with a reduced-parameter Geant4 optical model. The measurements establish the single-module photoelectron yield, longitudinal response, detection efficiency, and rate stability, while the model is constrained by the measured prototype responses and used to examine detector-design choices and momentum-dependent particle identification beyond the directly tested conditions. The quantitative PID study uses $K^+/p$ discrimination, and the level of data--simulation agreement relevant to these extrapolations is established in Sec.~\ref{subsec:calibration}.

\section{Detector concept and prototype design}
\label{sec:design}

\subsection{Threshold principle and detector requirements}

For a charged particle with velocity $\beta c$ traversing a medium of refractive index $n(\lambda)$, Cherenkov radiation is produced for $\beta n(\lambda)>1$ at the Cherenkov angle $\theta_C(\lambda)$
\begin{equation}
 \cos\theta_C(\lambda)=\frac{1}{\beta n(\lambda)}.
\end{equation}
For the E94 setting discussed above, the detector must respond to $K^+$ across the corresponding S-2S momentum acceptance while rejecting protons at the trigger level. More generally, the same radiator can be operated either as a classical threshold counter or, when both particle species are above threshold, as a yield-threshold device that exploits differences in the detected-NPE distributions. The longitudinal size of the active detector and the wide spatial distribution downstream of S-2S require efficient light collection over a long optical path in addition to sufficient local Cherenkov-photon production.

Table~\ref{tab:requirements} summarizes the principal detector requirements and the adopted geometry of one segment. We define the detector coordinates such that the nominal particle trajectory is along $z$, the S-2S dispersive direction is $x$, and the vertical direction is $y$. Here ``dispersive direction'' denotes the momentum-sensitive direction in which the downstream trajectory position changes predominantly with particle momentum after magnetic deflection. One radiator therefore measures \SI{730}{mm} in $y$, \SI{100}{mm} in $x$, and \SI{50}{mm} in $z$. The production S-2S detector consists of 16 such modules arranged side by side along the dispersive $x$ direction.

\begin{table*}[!t]
  \centering
  \caption{Principal design parameters and requirements for the S-2S acrylic Cherenkov counter. The J-PARC Main Ring operated with a \SI{4.24}{s} slow-extraction cycle; the quoted rate is the average extracted-particle rate during the approximately \SI{2}{s} beam-delivery portion of each cycle.}
  \label{tab:requirements}
  \small
  \begin{tabularx}{0.94\linewidth}{@{}p{0.42\linewidth}X@{}}
    \toprule
    Quantity & Requirement or adopted value \\
    \midrule
    E94 $K^+$ acceptance for cited setting ($p_{\rm cent}\simeq\SI{0.72}{GeV/\clight}$) & \SIrange{0.61}{0.83}{GeV/\clight} \\
    Acrylic refractive index & $n\simeq1.49$ \\
    Cherenkov threshold, $K^+$ & approximately \SI{0.45}{GeV/\clight} \\
    Cherenkov threshold, proton & approximately \SI{0.85}{GeV/\clight} \\
    Target $K^+$ detection efficiency & $>\SI{95}{\percent}$ at $T=1$ photoelectron \\
    Radiator size per segment ($y\times x\times z$) & \SI{730}{mm}$\times$\SI{100}{mm}$\times$\SI{50}{mm} \\
    Number of modules in S-2S & 16 (arrayed along $x$) \\
    Photosensor & Hamamatsu H11284-100UV, two-ended readout \\
    Beam-test average rate during beam delivery & approximately $2.5\times10^4$--$2.5\times10^5~\mathrm{s^{-1}}$ \\
    & (50--500 k particles/spill) \\
    \bottomrule
  \end{tabularx}
\end{table*}

\begin{figure}[width=\columnwidth,pos=!htbp,align=\centering]
  \includegraphics[width=0.86\columnwidth]{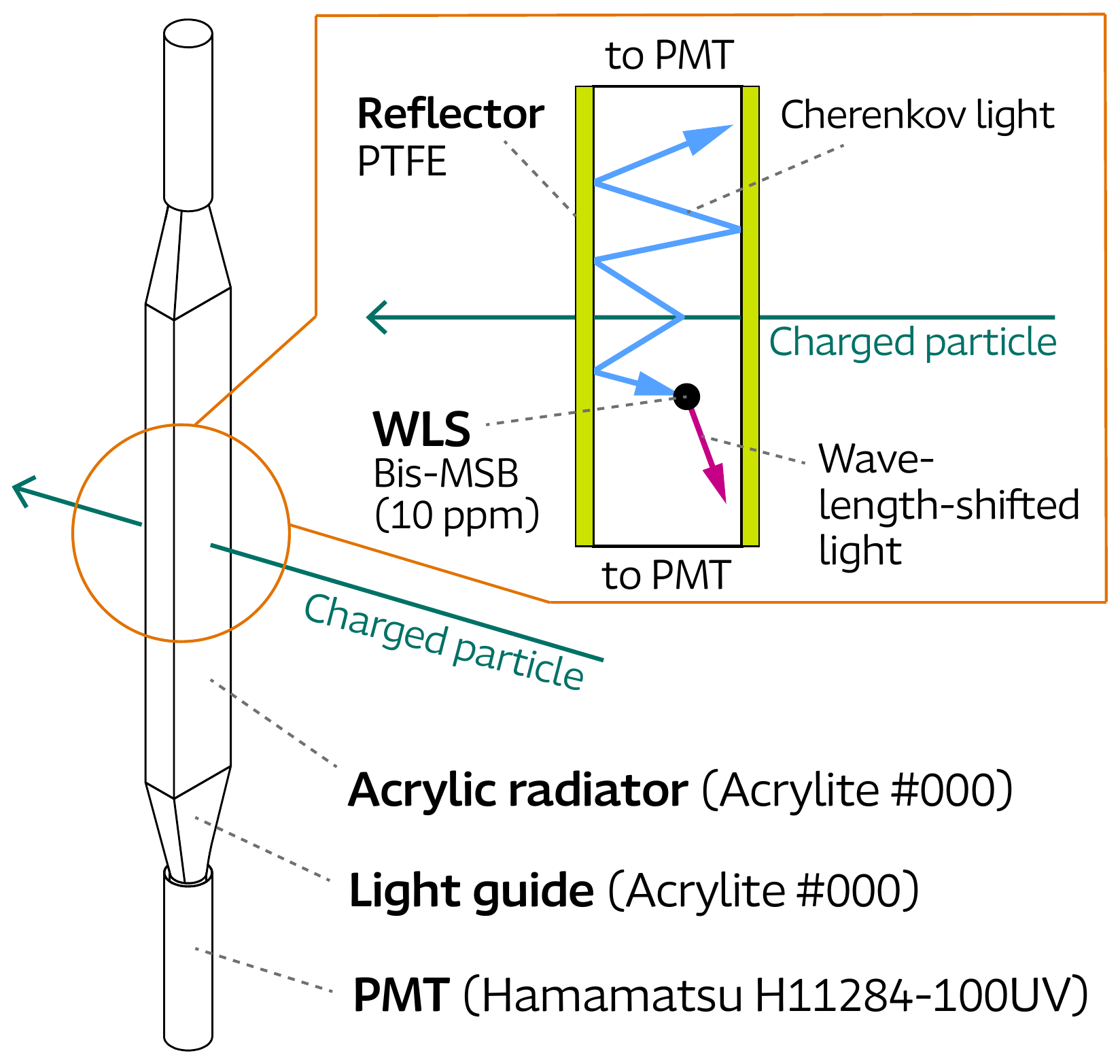}
  \caption{Conceptual layout of one UVT-acrylic Cherenkov-counter segment and the principal optical processes represented in the detector model. A charged particle traversing the Acrylite \#000 radiator produces Cherenkov photons that propagate toward the two photomultiplier readouts through acrylic light guides; the radiator is surrounded by a polytetrafluoroethylene (PTFE) reflector. Surface roughening provides one way of modifying the transport by redistributing photon directions at the acrylic boundary without changing wavelength. In the WLS-loaded configuration, bis-MSB at \SI{10}{ppm} by mass additionally absorbs short-wavelength photons and re-emits wavelength-shifted light with reduced memory of the original Cherenkov direction. The schematic illustrates the optical concept and is not drawn to scale.}
  \label{fig:concept}
\end{figure}

Figure~\ref{fig:concept} summarizes the detector concept and the two optical modifications compared in this study. In the polished/no-WLS configuration, Cherenkov photons are transported without spectral conversion and retain substantial memory of their initial emission and subsequent reflections. Surface roughening does not change the photon wavelength, but redistributes photon directions at the acrylic boundary and thereby changes the balance among trapped, reflector-coupled, and readout-directed trajectories. The bis-MSB-loaded configuration introduces a distinct bulk process in which short-wavelength photons are absorbed and re-emitted at longer wavelength with reduced directional memory.

\subsection{UV-transmitting acrylic and wavelength shifter}

The radiator material is Acrylite \#000 UV-transmitting acrylic. Its wavelength-dependent transmission was measured previously for \SI{5}{mm}-thick samples using a spectrophotometer \cite{Gogami2016WC}. The measured transmission extends to wavelengths substantially shorter than those transmitted by conventional acrylic and forms the experimental basis of the bulk-absorption model used here. In converting the measured sample transmittance to a bulk absorption length, the Fresnel losses at the two sample surfaces are removed so that surface reflection and bulk attenuation are treated separately in Geant4.

One prototype contains bis-MSB at \SI{10}{ppm} by mass. Bis-MSB is an aromatic fluorescent wavelength shifter with molecular formula $\mathrm{C}_{24}\mathrm{H}_{22}$ and molar mass \SI{310.44}{g/mol} \cite{RevvityBisMSB}. Its conjugated $\pi$-electron system absorbs short-wavelength light and re-emits at longer wavelength, with an emission maximum near \SI{415}{nm} whose exact position depends on the surrounding medium \cite{RevvityBisMSB}. This spectral conversion is advantageous because both acrylic transmission and PMT sensitivity are more favorable at longer wavelength. Quantitative absorption and fluorescence inputs for the optical model are based on spectroscopic measurements. In particular, the adopted fluorescence quantum yield is based on the measurement of Ding et al., $0.926\pm0.053$ for \SI{350}{nm} excitation in solution \cite{BisMSBOptics}. Because those measurements were not performed in the present poly(methyl methacrylate) (PMMA) matrix, the WLS absorption-length scale is treated as an effective model parameter constrained by the beam data. The \SI{10}{ppm} by-mass concentration is inherited from earlier detector experience and is not optimized in this study; the purpose of the comparison is to determine how wavelength shifting modifies the measured and simulated light-collection response relative to surface roughening.

\subsection{Mechanical and optical configuration}

Each prototype consists of a nominal \SI{730}{mm} ($y$)$\times$\SI{100}{mm} ($x$)$\times$\SI{50}{mm} ($z$) radiator read out at the two $y$ ends by Hamamatsu H11284-100UV PMTs through acrylic light guides. In S-2S operation, charged particles traverse approximately the \SI{50}{mm} thickness along $z$, while the 16 production modules tile the dispersive $x$ direction. For each configuration, the light guides were fabricated from the same acrylic formulation as the particle-illuminated radiator, including the presence or absence of bis-MSB, so that the optical material remained continuous in composition from the radiator into the light-guide region. The radiator and light-guide surfaces were wrapped with \SI{0.1}{mm}-thick polytetrafluoroethylene (PTFE) pipe-seal tape using half-width overlap, corresponding to approximately four effective layers. The PTFE was covered by one to two layers of \SI{12}{\micro m}-thick aluminum foil and an outer black light-tight sheet. In the beam-test assemblies, the acrylic/light-guide/PMT interfaces were coupled with EL-550 optical grease to permit disassembly. The production detector uses Dymax 3094/3099 UV-curing optical adhesive. The optical simulation represents the reflector and coupling structure at an effective level; uncertainties associated with the microscopic acrylic--air--PTFE contact condition are therefore treated as part of the boundary-model uncertainty discussed in Sec.~5.

The three beam-tested configurations were
\begin{enumerate}[label=(\arabic*)]
  \item polished UVT acrylic without bis-MSB (polished/no-WLS),
  \item UVT acrylic without bis-MSB with the four long surfaces roughened using \#400 sandpaper (rough/no-WLS), and
  \item polished UVT acrylic containing bis-MSB at \SI{10}{ppm} by mass (polished/WLS).
\end{enumerate}
The two end faces used for optical coupling remained polished in all three prototypes. This controlled set therefore separates surface-induced angular redistribution from the bulk wavelength-conversion and re-emission process introduced by bis-MSB.
Additional details of the detector development are given in Ref.~\cite{Taniguchi2026Thesis}.

\section{Hadron-beam test}
\label{sec:beamtest}

\subsection{Beam conditions and test setup}

The prototype counters were tested at the J-PARC K1.8 beam line in November 2025, with the three radiators installed downstream of S-2S. Beam-momentum settings were \SI{1.3}{GeV/\clight} and \SI{1.4}{GeV/\clight} for antiprotons and \SI{1.3}{GeV/\clight} for pions. The \SI{1.3}{GeV/\clight} antiproton sample is the principal data set for the NPE-response study and optical-model calibration. Its velocity is comparable to that of kaons near the low-momentum edge of the candidate E94 setting with $p_{\rm cent}\simeq\SI{0.72}{GeV/\clight}$, making it a relevant low-$\beta$ test condition for that setting. The pion runs sampled the radiator center and ends and were used to study longitudinal response and rate stability; the \SI{1.4}{GeV/\clight} antiproton data provide an additional kinematic check. The J-PARC Main Ring operated with a \SI{4.24}{s} slow-extraction cycle. Within each cycle, particles were delivered over an effective period of approximately \SI{2}{s}. The beam-test settings of 50, 200, and 500 k particles/spill therefore correspond to average extracted-particle rates during beam delivery of approximately $2.5\times10^4$, $1.0\times10^5$, and $2.5\times10^5~\mathrm{s^{-1}}$, respectively.

The S-2S magnetic system has a QQD configuration: two quadrupole magnets followed by one dipole magnet. For the present test, the beam-through optics kept the trajectories near the central axes of the two quadrupoles, while the dipole field setting was varied so that the particles did not in general follow the dipole central orbit. The resulting change in dipole deflection displaced the beam in the downstream detector region and provided the position scan used here. The existing S-2S detector package is located downstream of the dipole magnet; the prototype acrylic Cherenkov counters and a reference plastic scintillator were installed in the available downstream space, following the S-2S water Cherenkov detector~\cite{Gogami2016WC}.

The hit position was scanned along the \SI{730}{mm} long axis to measure the two single-ended responses and their sum. For the beam test, the prototypes were rotated so that this long axis lay along the direction in which the beam position moved when the S-2S dipole field setting was varied. This geometry converted the dipole-field scan into a longitudinal scan of the radiator. To keep the notation consistent with the installed S-2S detector, this long-axis coordinate is denoted $y$ throughout this paper; in the final detector, $y$ is vertical, $x$ is the S-2S dispersive direction, and particles traverse approximately along $z$. We define $y<0$ on the high-momentum side of the focal-plane acceptance, with corresponding track angle $\theta_y$. The readout channels are labeled by their physical ends, PMT($-y$) and PMT($+y$).

\begin{figure}[width=\columnwidth,pos=!htbp,align=\centering]
  \includegraphics[width=0.90\columnwidth]{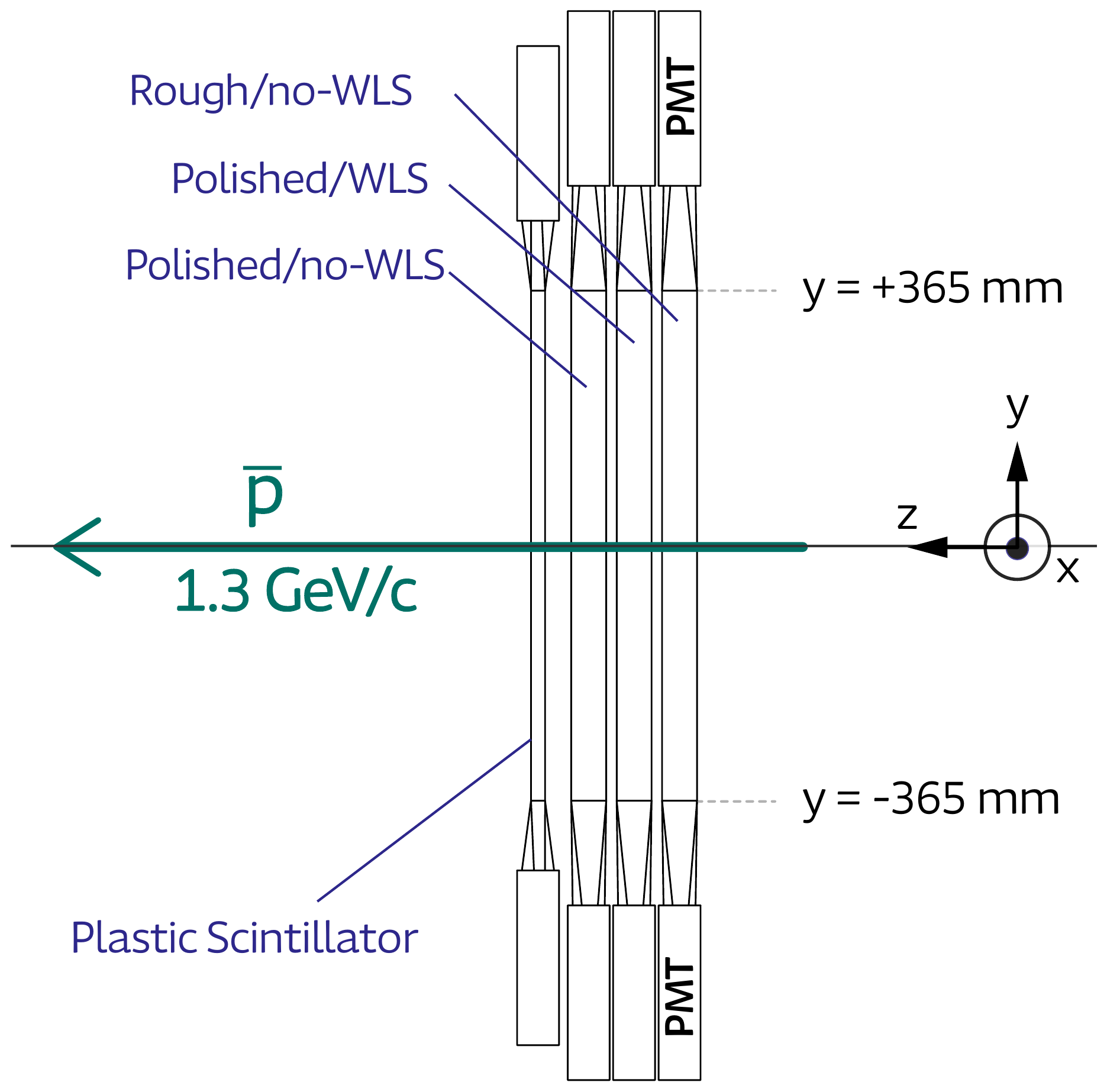}
  \caption{Beam-test setup for the three UVT-acrylic Cherenkov-counter prototypes. The rough/no-WLS, polished/WLS, and polished/no-WLS radiators were installed sequentially along the beam line and read out at both longitudinal ends by PMTs. The \SI{1.3}{GeV/\clight} antiproton beam propagated along the $z$ direction, while the long radiator axis defines the longitudinal coordinate $y$; the nominal radiator ends correspond to $y=\pm\SI{365}{mm}$. A downstream plastic scintillator provided an independent charged-particle reference used in the event selection.}
  \label{fig:beam-setup}
\end{figure}

Changing the dipole field setting varied the hit position and incidence angle together. Because the quadrupole transport remained close to the beam-through central axes while the trajectory through the dipole was intentionally displaced, the observed position--angle correlation was governed predominantly by the dipole deflection. The measured conditional distribution $P(\theta_y|y)$ is used directly to define the incident phase space for the Geant4 comparison. Figure~\ref{fig:beam-setup} shows the relative placement of the three prototypes, the two-ended readout, the downstream reference scintillator, and the coordinate convention used throughout the paper.

\begin{figure*}[width=\textwidth,pos=!t,align=\centering]
  \centering
  \includegraphics[width=0.98\linewidth]{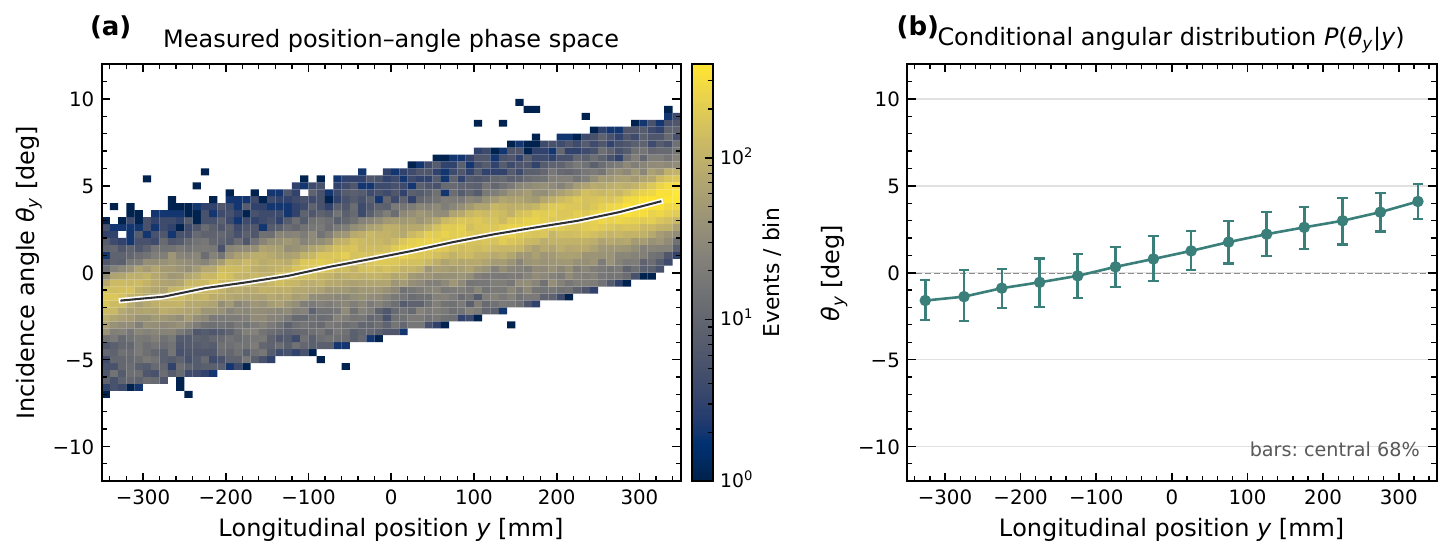}
  \caption{Measured longitudinal position--angle phase space for the \SI{1.3}{GeV/\clight} antiproton sample. (a) Event density as a function of reconstructed longitudinal hit position $y$ and incidence angle $\theta_y$; the overlaid curve shows the mean angular trend in longitudinal-position bins. (b) Conditional angular response summarized by the mean $\theta_y$ in each $y$ bin. Vertical bars indicate the central 68\% of the measured angular distribution rather than the statistical uncertainty on the mean. The correlation is produced by the dipole-setting scan used to vary the beam position along the radiator. The measured $P(\theta_y|y)$ is used to define the incident phase space in Geant4.}
  \label{fig:phase-space}
\end{figure*}

Figure~\ref{fig:phase-space} shows the measured longitudinal position--angle phase space for the selected \SI{1.3}{GeV/\clight} antiproton sample. Both the mean angle and its finite spread vary with longitudinal position. The response at a given $y$ therefore averages over the measured conditional distribution $P(\theta_y|y)$, which is imposed in the Geant4 baseline before any optical parameters are varied.

\subsection{Readout electronics and trigger}

Each PMT signal was split into charge and timing branches. The charge signal was digitized with a CAEN V792 charge-to-digital converter (QDC), while timing was recorded with a low-resolution multi-hit time-to-digital converter (TDC) implemented on a Hadron Universal Logic (HUL) FPGA module \cite{Honda2021HUL}. The beam-test trigger required a coincidence of the K1.8 beam-line hodoscopes BH1 and BH2 \cite{Takahashi2012K18} with the downstream S-2S time-of-flight (TOF) scintillator wall \cite{Gogami2016WC}, selecting charged particles transported to the downstream detector region. The two PMT charges from each prototype were recorded independently and combined event by event in the analysis.

\subsection{Photoelectron calibration and event selection}

The charge recorded in each PMT was converted to a photoelectron count using the measured pedestal and single-photoelectron charge separation,
\begin{equation}
 N_{\rm PE}=\frac{Q-Q_{\rm ped}}{\Delta Q_{1\rm PE}},
\end{equation}
where $Q$ is the event-by-event digitized charge, $Q_{\rm ped}$ is the pedestal position, and $\Delta Q_{1\rm PE}$ is the measured charge separation between the pedestal and single-photoelectron peaks. The two calibrated ends were then summed event by event,
\begin{equation}
 N_{\rm PE}^{\rm SUM}=N_{\rm PE}^{(-y)}+N_{\rm PE}^{(+y)}.
\end{equation}
This event-by-event quantity is referred to as the software SUM. For the longitudinal response, a local Gaussian-plus-constant fit is used in each position bin; its fitted peak position is referred to as the MPV estimator. Statistical uncertainties on these peak values are evaluated by bootstrap resampling.

Charged-particle trajectories were reconstructed with the S-2S multi-wire drift chambers, whose sub-millimeter position resolution is negligible on the scale of the \SI{50}{mm} longitudinal bins used here. Antiproton candidates are required to satisfy the reference-scintillator condition and to contain a single reconstructed S-2S trajectory with $\chi^2/\mathrm{ndf}<2$. The efficiency analysis further uses $|y|<\SI{350}{mm}$ and $|x|<\SI{20}{mm}$. These selections define the efficiency denominator independently of the Cherenkov-counter response.

\subsection{Efficiency definition}

The detection efficiency is defined from the event-by-event software SUM as
\begin{equation}
\begin{aligned}
 \epsilon(T) &= \frac{N_{\rm hit}(T)}{N_{\rm selected}}, \\
 N_{\rm hit}(T) &\equiv
 N\!\left[N_{\rm PE}^{\rm SUM}>T\,;\,\text{selected sample}\right].
\end{aligned}
\end{equation}
Here $N_{\rm selected}$ is the number of events passing the independent beam and tracking selections, and $N_{\rm hit}(T)$ is the subset of those events whose software-SUM response exceeds the threshold $T$. Unless otherwise stated, the nominal threshold is $T=1$ photoelectron, and statistical intervals are evaluated as binomial confidence intervals. This software SUM reproduces the intended segment-level trigger quantity, for which the two PMT signals are analog-summed before discrimination. Because the prototype data set recorded the two ends independently, the efficiency reported here is calculated from the calibrated QDC responses. A subsequent beam test using the actual analog-sum circuit gave consistent results, validating this offline procedure.

\section{Beam-test results}
\label{sec:results}

\subsection{Photoelectron yield and longitudinal response}

The three prototypes exhibit clearly different longitudinal responses. The polished/no-WLS radiator gives the lowest photoelectron yield. Roughening the four long surfaces increases the collected light and changes the longitudinal response. The polished/WLS radiator gives the largest two-ended yield over most of the scanned range. In all configurations the individual PMT responses vary with position, whereas their event-by-event sum compensates a substantial part of this variation.

\begin{figure}[width=\columnwidth,pos=!htbp,align=\centering]
  \includegraphics[width=\columnwidth]{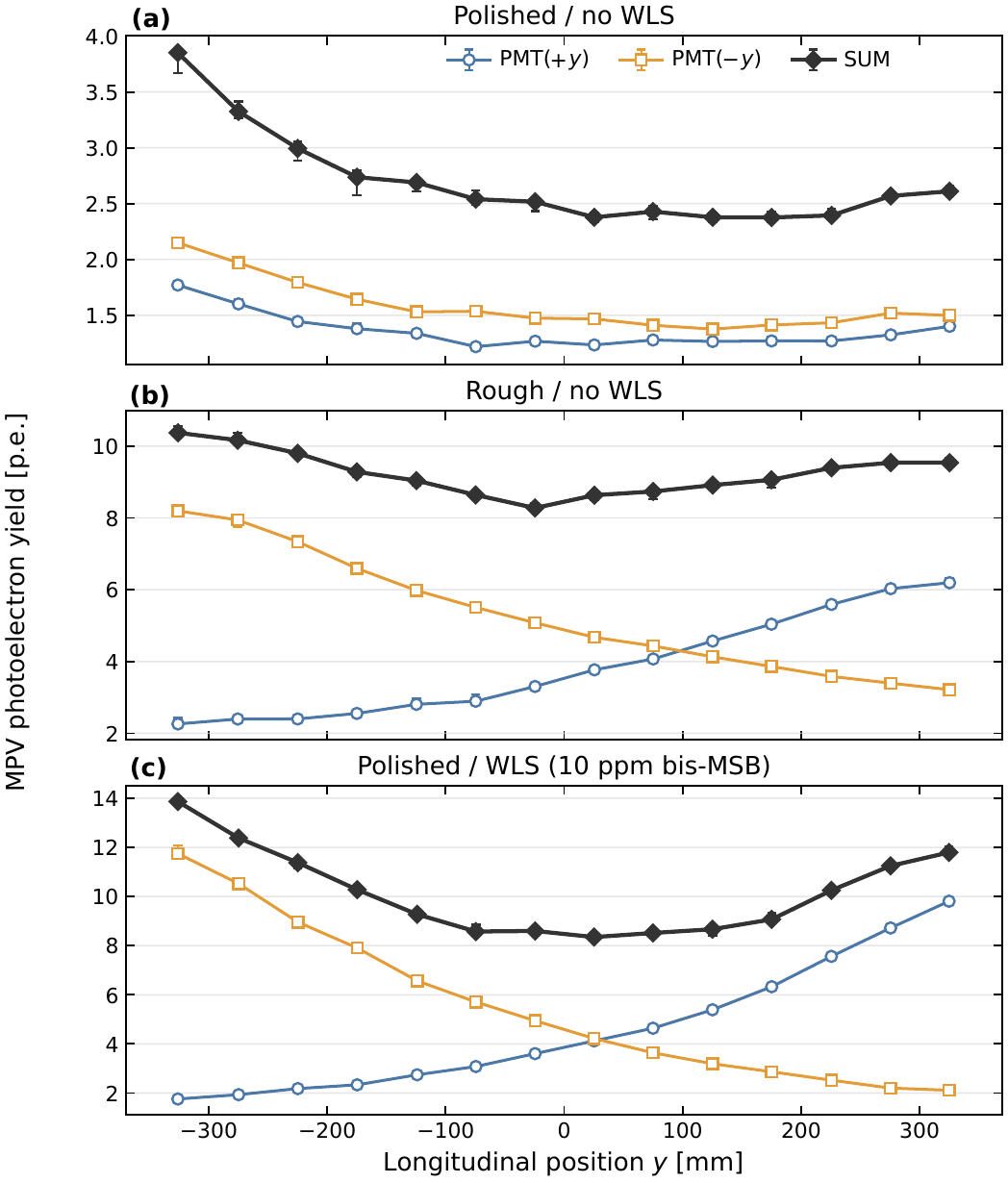}
  \caption{MPV estimator of the photoelectron yield as a function of longitudinal position for (a) the polished/no-WLS, (b) rough/no-WLS, and (c) polished/WLS (bis-MSB at \SI{10}{ppm} by mass) prototypes in the \SI{1.3}{GeV/\clight} antiproton data. The PMT($-y$), PMT($+y$), and event-by-event SUM responses are shown for each configuration. Error bars represent the statistical uncertainty of the MPV estimator. The longitudinal coordinate follows the common convention defined in Sec.~\ref{sec:beamtest}.}
  \label{fig:npe-position}
\end{figure}

Roughening and wavelength shifting alter both the collected light and its longitudinal dependence, but in different ways. The rough/no-WLS configuration collects more light than the polished/no-WLS reference, while the polished/WLS configuration gives the largest SUM response over most of the scanned region. The two single-ended responses vary in opposite directions with position: approaching one end increases the nearby-PMT response while decreasing the far-side response, so two-ended summing reduces the net longitudinal dependence. The single-photoelectron charge calibration corrects the charge scale of each PMT but does not equalize intrinsic photon-detection efficiency among tubes. Previous measurements with the same H11284-100UV model found an approximately 6\% tube-to-tube standard deviation in detected NPE \cite{Gogami2016WC}. Part of the absolute difference between the two single-ended response levels can therefore arise from PMT variation as well as optical transport.

\begin{figure*}[width=\textwidth,pos=!t,align=\centering]
  \centering
  \includegraphics[width=0.98\linewidth]{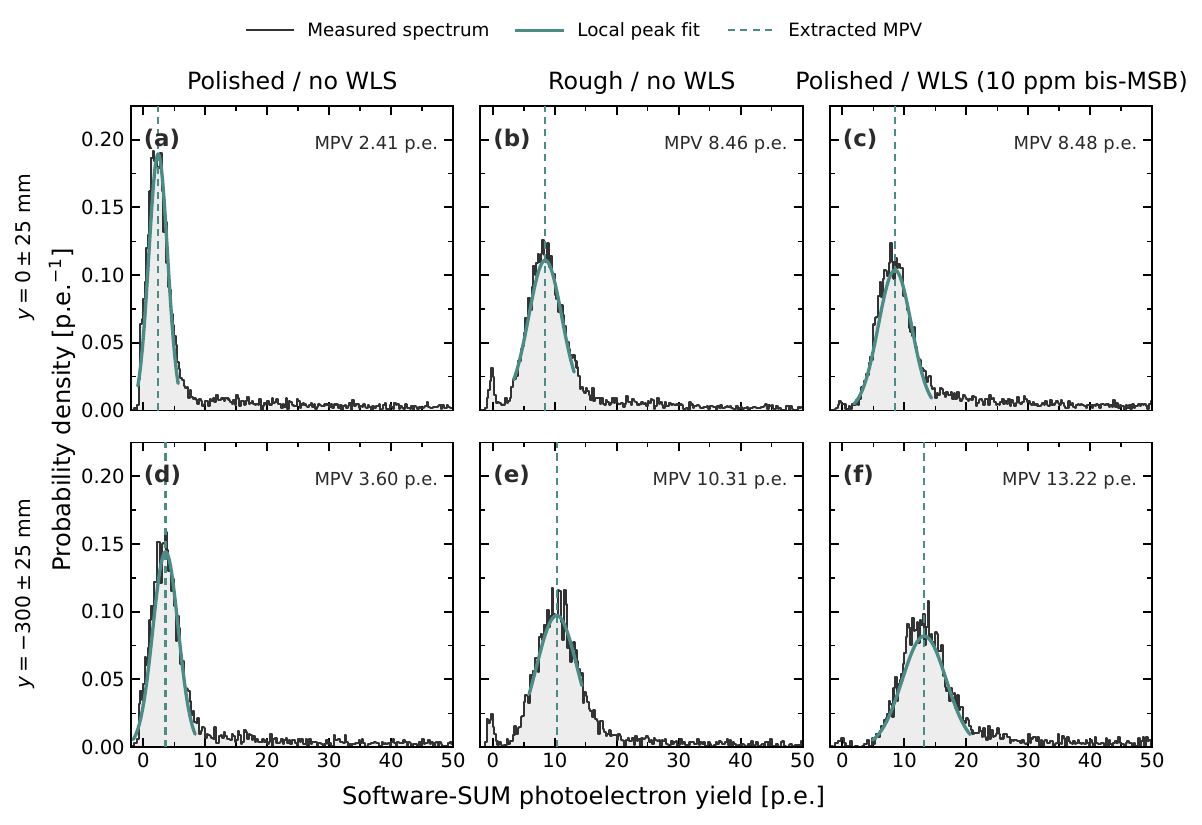}
  \caption{Representative software-SUM photoelectron distributions measured for the three prototype configurations at two longitudinal positions in the \SI{1.3}{GeV/\clight} antiproton data. Panels (a)--(c) correspond to $y=0\pm\SI{25}{mm}$ and panels (d)--(f) to $y=-300\pm\SI{25}{mm}$ for the polished/no-WLS, rough/no-WLS, and polished/WLS configurations, respectively. The black histograms show the measured distributions, the teal curves indicate the local Gaussian-plus-constant fits used to determine the peak position, and the dashed vertical lines mark the corresponding MPV estimators. The full distributions are shown to illustrate the configuration- and position-dependent widths and non-Gaussian high-photoelectron tails in addition to the peak response.}
  \label{fig:npe-spectrum}
\end{figure*}

Figure~\ref{fig:npe-spectrum} shows software-SUM NPE spectra at the radiator center and near one end. In addition to the fitted peak position, the distributions exhibit configuration- and position-dependent widths and non-Gaussian high-NPE tails. These features are not required for the efficiency measurement, but they provide additional distribution-level information for the optical-model comparison.

\subsection{Detection efficiency and rate stability}

\begin{figure}[width=\columnwidth,pos=!htbp,align=\centering]
  \includegraphics[width=0.86\columnwidth]{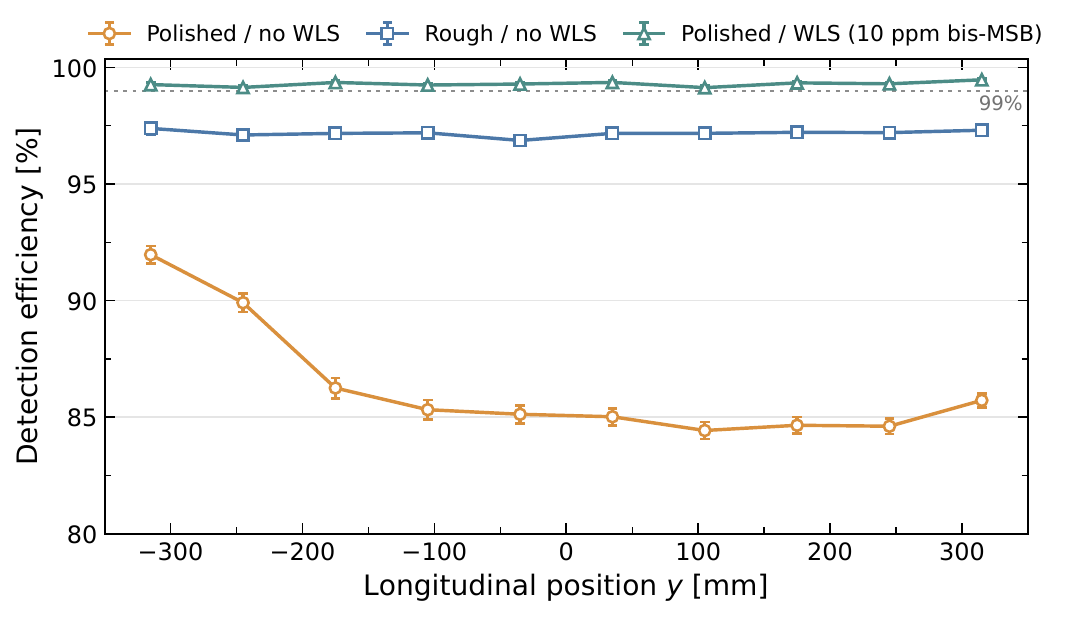}
  \caption{Longitudinal dependence of the software-SUM detection efficiency measured with the \SI{1.3}{GeV/\clight} antiproton beam at the nominal threshold of $T=1$ photoelectron. Error bars indicate 68\% Wilson intervals. The polished/WLS prototype maintains a detection efficiency above \SI{99}{\percent} throughout the scanned longitudinal region.}
  \label{fig:efficiency}
\end{figure}

Figure~\ref{fig:efficiency} summarizes the longitudinal dependence of the operational software-SUM efficiency at the nominal discriminator-equivalent threshold of $T=1$ photoelectron. The polished/WLS prototype remains above \SI{99}{\percent} throughout the scanned region, with values of approximately \SI{99.1}{\percent}--\SI{99.5}{\percent}. The rough/no-WLS efficiency is approximately \SI{97}{\percent}, while the polished/no-WLS efficiency is lower and more position dependent. The two-ended sum therefore maintains high efficiency even though the individual PMT responses vary with position.

Rate stability was checked at the radiator center with a \SI{1.3}{GeV/\clight} pion beam over average extracted-particle rates during beam delivery of approximately $2.5\times10^4$--$2.5\times10^5~\mathrm{s^{-1}}$ (50--500 k particles/spill). No significant dependence of the two-ended efficiency on rate was observed. At all three tested rate settings, the software-SUM efficiency remained above 99\%, consistent with the high efficiency observed in the longitudinal scan.

\section{Data-constrained optical model}
\label{sec:optical-model}

\subsection{Role and scope of the model}

A Geant4 optical model of one detector segment is used to interpret the measured response in terms of effective detector optical properties and light-collection behavior \cite{Agostinelli2003Geant4,Allison2006Geant4,Allison2016Geant4}. Independently measured or externally constrained inputs are fixed where possible; four detector-level optical parameters that remain uncertain in the assembled counter are varied over physically motivated ranges and constrained by the beam data. The calibration therefore tests whether one common parameter set can account for the relative response scale and spatial behavior of configurations with different surface treatment and wavelength shifting.

The calibrated model is then checked against longitudinal observables and the polished/no-WLS transfer case before its parameters are held fixed for detector-design comparisons and momentum-dependent $K^+/p$ predictions. Section~\ref{subsec:calibration} quantifies the data--simulation agreement and identifies the aspects of the optical description that remain model dependent.

\subsection{Optical model}

The Geant4 model contains the measured wavelength-dependent transmission of Acrylite \#000, converted to a bulk attenuation description, together with the refractive index, acrylic--air and air--PTFE boundaries, PTFE reflectivity, optical coupling to the light guides and PMTs, and the PMT quantum efficiency. Polished and rough acrylic boundaries are represented with the Geant4 UNIFIED surface model, with the rough-surface parameter treated as an effective parameterization of the \#400 finish.

For the WLS-loaded radiator, bis-MSB absorption and re-emission are added to the same underlying UVT-acrylic description. The spectral shapes are taken from published optical data, and the fluorescence quantum yield is set to 0.93, consistent with the measured value reported for \SI{350}{nm} excitation \cite{BisMSBOptics}. Because the relevant spectroscopy was not measured in the present PMMA matrix, the WLS absorption-length scale is retained as an effective detector-level parameter constrained by the beam data. Surface roughening and wavelength shifting are therefore represented as physically distinct modifications of the optical response: the former alters boundary-induced angular redistribution, whereas the latter introduces bulk spectral conversion and re-emission.

\subsection{Calibration and validation strategy}
\label{subsec:calibration}

Inputs known independently of the beam test are fixed wherever possible. The remaining detector-level uncertainty is represented by four effective optical parameters,
\begin{equation}
 \boldsymbol{\theta}=(R_{\rm PTFE},A_{\rm WLS},\sigma_{\alpha,\rm rough},S_{\rm UVT}),
\end{equation}
Here $R_{\rm PTFE}$ is a multiplicative scale applied to the reference PTFE reflectivity spectrum, $A_{\rm WLS}$ scales the WLS absorption-length spectrum, $\sigma_{\alpha,\rm rough}$ is the effective Geant4 UNIFIED-model surface-roughness parameter used to represent the \#400 finish, and $S_{\rm UVT}$ scales the Acrylite \#000 bulk absorption-length spectrum. A separate factor $S_{\rm resp}$ is applied at the analysis stage as one common multiplicative scale on the simulated NPE axis, shared by the calibration configurations, PMTs, and hit positions. Thus the four optical parameters control relative optical response, while $S_{\rm resp}$ accounts only for a common overall detector-response scale. No PMT-specific or prototype-specific response factor is introduced. The channel-by-channel single-photoelectron calibration removes charge-gain differences at first order, but intrinsic tube-to-tube photon-detection differences remain in the measured NPE and therefore contribute to the data--simulation residuals rather than being tuned away. The approximately 6\% PMT-to-PMT spread observed previously for the same H11284-100UV model \cite{Gogami2016WC} provides a useful reference scale for possible PMT-to-PMT contributions to the observed residuals.

The primary calibration uses the rough/no-WLS and polished/WLS prototypes, whose responses provide complementary constraints on surface and WLS effects. Event-level software-SUM NPE distributions in the calibration region are compared after applying the same common response scale to both configurations. For each optical-parameter tuple, $S_{\rm resp}$ is profiled by minimizing the mean Jensen--Shannon divergence. Defining $\mathcal{C}=\{\text{rough/no-WLS},\,\text{polished/WLS}\}$, the objective is
\begin{equation}
\begin{aligned}
 {\cal J}(\boldsymbol{\theta})
 &= \min_{S_{\rm resp}}\;\frac{1}{2}
 \sum_{c\in\mathcal{C}} \\
 &\quad\times D_{\rm JS}\!\left[
 f_c^{\rm data},f_c^{\rm MC}(S_{\rm resp};\boldsymbol{\theta})\right].
\end{aligned}
\end{equation}
Here $c$ labels the prototype configuration, $f_c^{\rm data}$ is the normalized measured software-SUM NPE distribution, and $f_c^{\rm MC}$ is the corresponding normalized simulated distribution after application of the common response scale. $D_{\rm JS}$ denotes the Jensen--Shannon divergence.
The nominal Rank~1 solution is selected from this ranking and its parameters are fixed for subsequent predictions; nearby Rank~2 and Rank~3 solutions are retained as alternative calibration solutions for a local robustness check.

The polished/no-WLS prototype serves as an additional transfer test of the same fixed parameter set. It exhibits a larger zero-response component in the simulation than in the beam data, identifying a limitation of the effective polished-boundary description. This mismatch limits quantitative predictions for polished/no-WLS but does not directly enter the momentum-dependent PID result, which uses the calibrated polished/WLS model at the radiator center.

\begin{figure*}[width=\textwidth,pos=!t,align=\centering]
  \centering
  \includegraphics[width=0.98\linewidth]{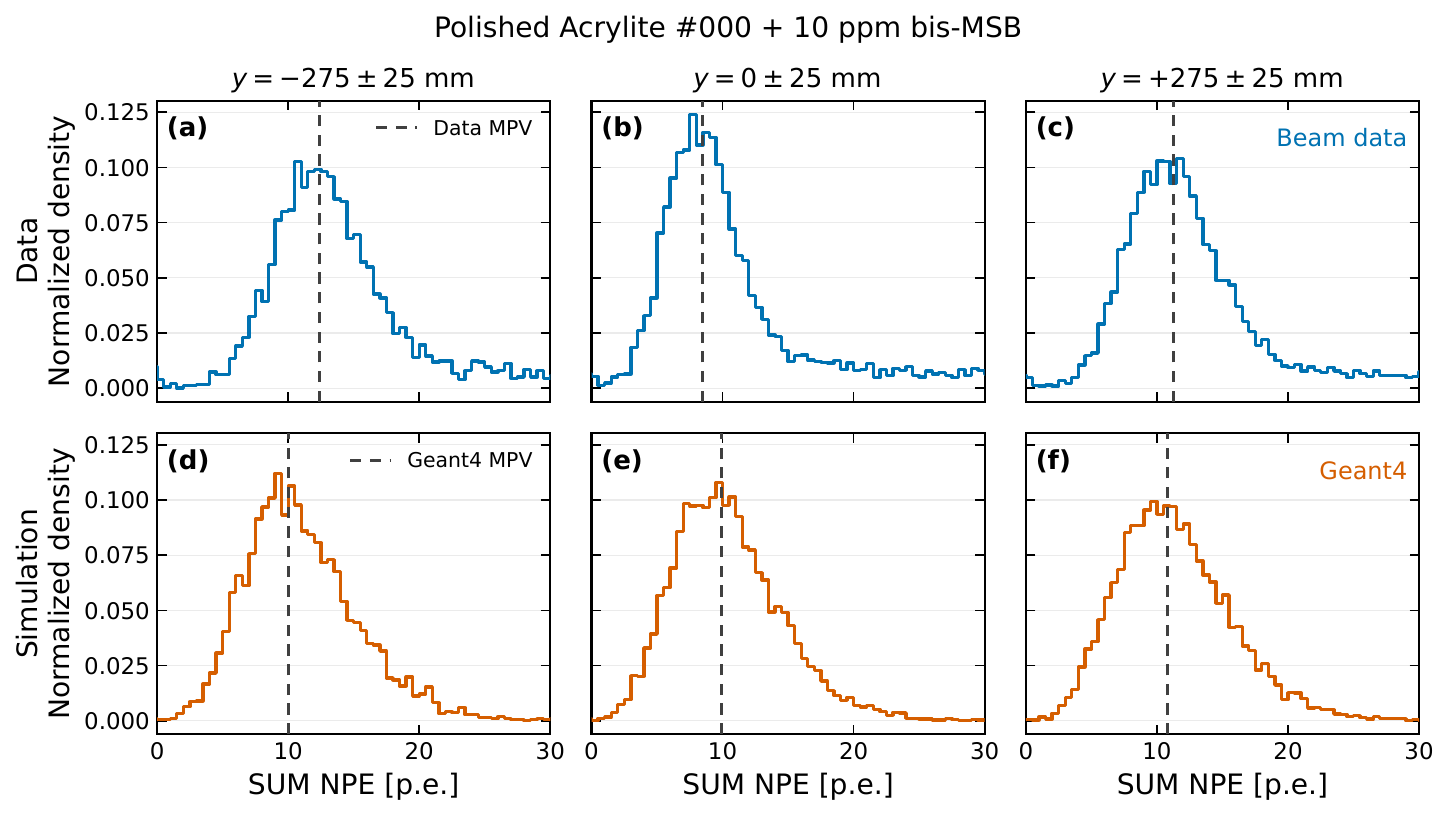}
  \caption{Representative software-SUM photoelectron distributions for the polished Acrylite \#000 radiator containing bis-MSB at \SI{10}{ppm} by mass. The upper row shows the \SI{1.3}{GeV/\clight} antiproton data and the lower row the calibrated Geant4 simulation at $y=-275\pm\SI{25}{mm}$, $0\pm\SI{25}{mm}$, and $+275\pm\SI{25}{mm}$. Histograms are shown with 0.5-p.e. display binning. Dashed vertical lines mark the MPV estimators obtained by applying the same local Gaussian-plus-constant peak-fit procedure to the measured and simulated distributions; the fit functions themselves are omitted for clarity. The optical parameters and common detector-response scale are fixed before the simulation comparison.}
  \label{fig:template-best}
\end{figure*}

After calibration, the longitudinal response is evaluated with the same local Gaussian-plus-constant peak estimator used for the beam data. Figure~\ref{fig:template-best} compares polished/WLS SUM-NPE distributions at three longitudinal positions. The same optical parameters and common response normalization are used throughout, so the comparison tests the transfer of the calibrated response across the radiator.

\begin{figure*}[width=\textwidth,pos=!t,align=\centering]
  \centering
  \includegraphics[width=0.98\linewidth]{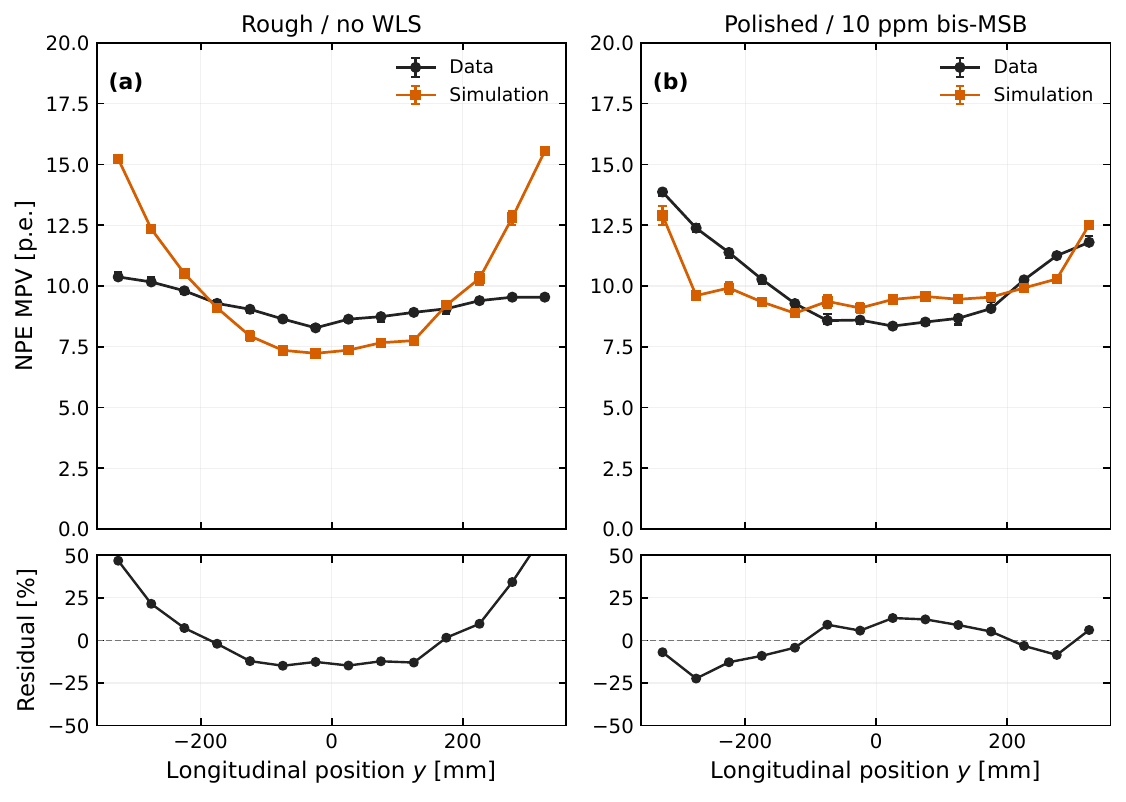}
  \caption{Longitudinal dependence of the software-SUM NPE MPV estimator for the rough/no-WLS and polished/WLS configurations after the optical parameters and common detector-response scale have been fixed. Data and simulation use the same local peak estimator in corresponding \SI{50}{mm}-wide longitudinal bins. The upper panels show the MPV estimators and the lower panels the fractional simulation--data residuals.}
  \label{fig:y-validation}
\end{figure*}

Figure~\ref{fig:y-validation} compares the corresponding longitudinal MPV estimators for the two configurations entering the primary calibration. Over the central $|y|<\SI{200}{mm}$ region, spanning \SI{400}{mm} or about 55\% of the \SI{730}{mm} radiator length, the simulated MPV estimators typically agree with the measured values at approximately the 15\% level. Larger structured deviations occur toward the ends, most clearly because the simulated rough/no-WLS response varies more strongly with position than observed. Accordingly, extrapolations that depend on rough-surface behavior near the radiator ends are treated qualitatively or semi-quantitatively. The $K^+/p$ study instead uses polished/WLS at the radiator center, within the region where the model-data agreement is characterized above.

\section{Predictions with fixed model parameters}
\label{sec:predictions}

The optical parameters and the common detector-response normalization used in this section are fixed by the calibration and validation procedure of Sec.~\ref{sec:optical-model}. The same fixed parameter set is used below to examine detector choices and kinematic conditions beyond the beam-test observables.

\subsection{Fixed-parameter detector-design comparison}
\label{subsec:designpred}

Once the optical parameters and common response scale are fixed by the template-based procedure described above, the model can be used to compare detector configurations that were not included in the beam test. The rough/no-WLS and polished/WLS prototypes provide the primary calibration constraints, while polished/no-WLS remains an explicit out-of-fit validation case. These model calculations use the same fixed parameter set with a central, normal-incidence reference trajectory at \SI{1.3}{GeV/\clight}. This construction isolates design choices on a common calibrated basis.

The fixed-parameter comparison focuses on three WLS-loaded configurations: a polished Acrylite \#001 + WLS reference representing a conventional, non-UVT acrylic baseline, the adopted polished Acrylite \#000 UVT + WLS design, and an unmeasured rough Acrylite \#000 UVT + WLS configuration. The \#001 entry provides the conventional-acrylic comparison. Table~\ref{tab:design-counterfactual} compares the event-averaged total NPE, including zero-photoelectron events, and normalizes each result to the polished Acrylite \#000 UVT + WLS response at $y=0$. The mean is used here because the predicted spectra can be strongly non-Gaussian and can contain a substantial near-zero component, for which a common peak estimator is not robust across configurations.

\begin{table*}[!t]
  \centering
  \caption{Fixed-parameter normal-incidence design comparison at \SI{1.3}{GeV/\clight}. The mean NPE includes zero-photoelectron events. ``Relative'' values are normalized to the polished Acrylite \#000 UVT + WLS mean response at $y=0$. The \#001 + WLS entry is the conventional-acrylic comparison. All rows use the same fixed optical model and common detector-response scale. Monte Carlo statistical errors on the quoted means are below 0.02 photoelectrons and are small compared with the model dependence discussed in the text.}
  \label{tab:design-counterfactual}
  \small
  \setlength{\tabcolsep}{4pt}
  \begin{tabular}{lccccc}
    \toprule
    Configuration & $\langle N_{\rm PE}\rangle_{0}$ & Rel.$_{0}$ & $\langle N_{\rm PE}\rangle_{300}$ & Rel.$_{300}$ & 300/0 \\
    \midrule
    Acrylite \#001, polished/WLS       & 1.25 & 0.12 & 1.77 & 0.17 & 1.42 \\
    Acrylite \#000 UVT, polished/WLS  & 10.44 & 1.00 & 11.32 & 1.08 & 1.08 \\
    Acrylite \#000 UVT, rough/WLS     & 0.92 & 0.09 & 11.18 & 1.07 & 12.2 \\
    \bottomrule
  \end{tabular}
\end{table*}

The fixed model predicts a large material dependence: at the radiator center, the Acrylite \#001/WLS response is about 12\% of that of the polished UVT/WLS reference. The unmeasured rough+WLS configuration behaves differently. Its central mean response is only about 9\% of the polished UVT/WLS value, but near $y=300$~mm it rises to essentially the same mean response as the polished UVT/WLS configuration. Thus roughening does not provide a uniform additional light-yield gain when combined with wavelength shifting in this model; instead, it produces a very strong redistribution of the longitudinal response.

The magnitude of this rough+WLS position dependence must be interpreted cautiously. Figure~\ref{fig:y-validation} already shows that the calibrated simulation overpredicts the measured longitudinal variation of the rough/no-WLS prototype. The rough+WLS calculation therefore inherits a known sensitivity of the effective rough-surface description, and its 300/0 ratio should not be read as a precision prediction for an unbuilt detector. Its qualitative message is nevertheless useful: with the same fixed parameter set, introducing rough boundaries into the WLS-loaded UVT radiator substantially changes the predicted longitudinal light-collection pattern and can strongly suppress the central response while preserving or enhancing end-region collection.

\subsection{Momentum-dependent $K^+/p$ discrimination}
\label{subsec:pid-scope}

The calibrated polished/WLS model is also used to estimate the momentum range over which the counter can select kaons while suppressing protons. The calculation uses a simple reference geometry: particles are generated at the radiator center, $y=0$, at normal incidence with zero angular spread. This geometry removes the position--angle correlation specific to the prototype beam test and avoids the radiator ends, where the optical response is less tightly constrained. Because the polished/WLS light yield is relatively low near the center, the chosen trajectory is conservative with respect to kaon light yield, although not necessarily with respect to proton rejection over an arbitrary experimental acceptance.

At each incident momentum, separate $K^+$ and proton samples are generated with identical momentum and detector geometry. For a chosen target kaon survival $\epsilon_K$, the required software-SUM threshold $T_{\rm req}(p)$ is determined from the simulated kaon NPE distribution through
\begin{equation}
 P(N_{\rm PE}^{K}>T_{\rm req}(p))=\epsilon_K,
\end{equation}
The threshold used to evaluate the proton survival is the applied threshold $T_{\rm applied}(p)$ defined below in Eq.~(10),
\begin{equation}
 \epsilon_p=P(N_{\rm PE}^{p}>T_{\rm applied}(p)).
\end{equation}
The nominal operating points considered are $\epsilon_K=0.99$, 0.90, and 0.80. Survival probabilities are evaluated over the full simulated event sample for each species and momentum. Below the kaon Cherenkov threshold this construction is not physically useful because the selected species itself does not provide a Cherenkov signal; those momenta are therefore excluded from the quantitative PID range.

\begin{figure*}[width=\textwidth,pos=!t,align=\centering]
  \centering
  \includegraphics[width=0.98\linewidth]{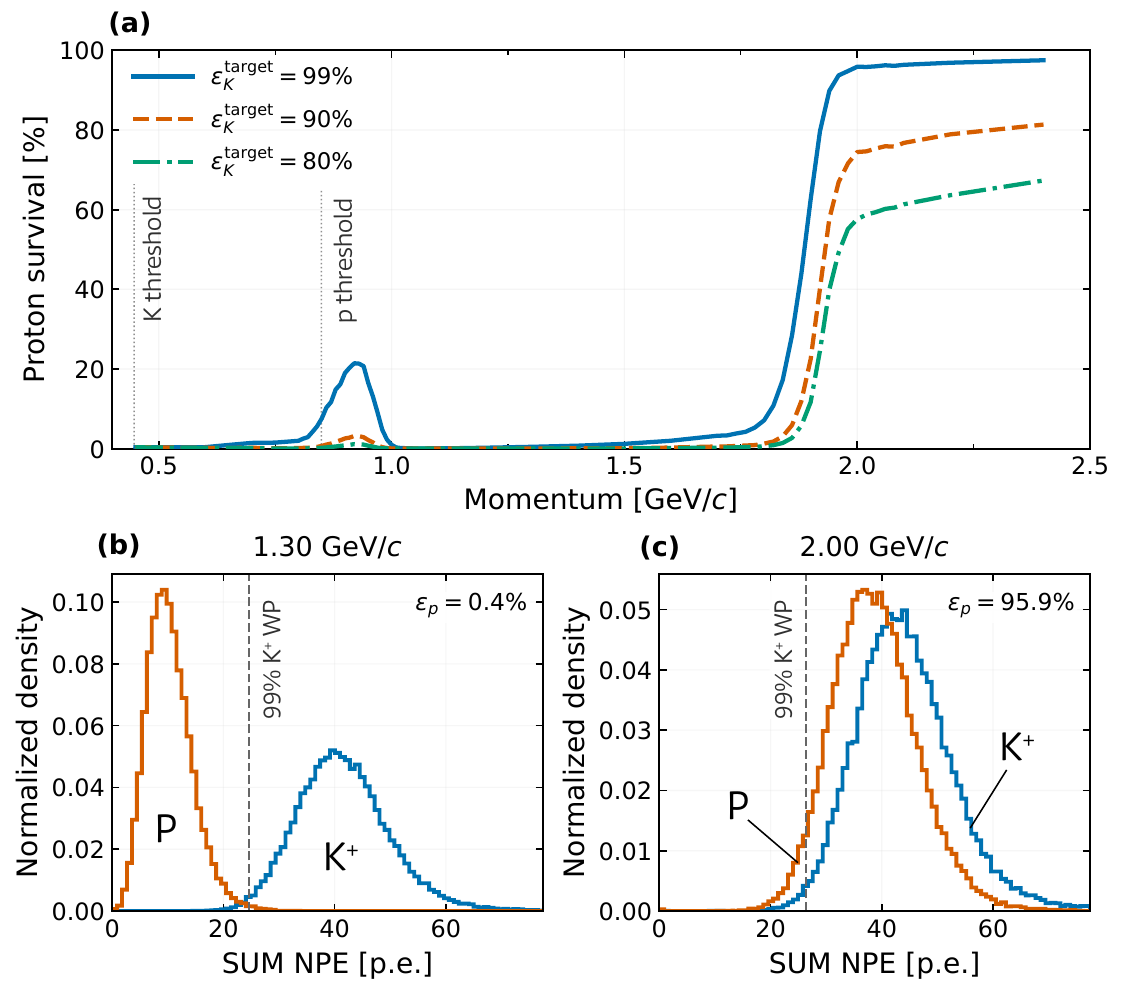}
  \caption{Predicted $K^+/p$ discrimination for the calibrated polished/WLS counter in the central, normal-incidence reference geometry. (a) Proton survival $\epsilon_p$ as a function of momentum for target $K^+$ survivals of 99\%, 90\%, and 80\%; the vertical dotted lines indicate the nominal Cherenkov thresholds for $K^+$ and protons in acrylic. (b,c) Normalized SUM-NPE distributions for $K^+$ and protons at \SI{1.30}{GeV/\clight} and \SI{2.00}{GeV/\clight}, respectively. The dashed vertical lines show the discriminator corresponding to the 99\% kaon-survival working point. The predicted proton survivals at these two momenta are 0.4\% and 95.9\%, illustrating the evolution from strong separation to substantial overlap.}
  \label{fig:pid-momentum-scope}
\end{figure*}

The beam data show that the polished/WLS prototype already achieves high detection efficiency at a software-SUM threshold of 1 photoelectron. The simulated discriminator is therefore subject to the practical floor
\begin{equation}
 T_{\rm applied}(p)=\max\!\left[T_{\rm req}(p),1~{\rm p.e.}\right].
\end{equation}
When this floor is active, the achieved kaon survival differs from the nominal target and is evaluated explicitly from the simulated sample. In Fig.~\ref{fig:pid-momentum-scope}, this occurs only at the lowest-momentum points near the $K^+$ Cherenkov threshold (approximately \SIrange{0.4}{0.5}{GeV/\clight}). The floor is inactive over the higher-momentum region used below to characterize the upper reach of the $K^+/p$ separation.

Figure~\ref{fig:pid-momentum-scope} summarizes the predicted proton survival as a function of momentum for the three nominal kaon-survival targets. The same figure shows $K^+$ and proton SUM-NPE distributions at \SI{1.30}{GeV/\clight} and \SI{2.00}{GeV/\clight} for the 99\% kaon-survival working point, illustrating the transition from strong separation to substantial overlap. At high momentum, loss of PID power is identified by $\epsilon_p\rightarrow\epsilon_K$; in this limit the two species become indistinguishable for the chosen one-dimensional NPE cut. In the central normal-incidence reference geometry, the fine momentum scan predicts that 90\% $K^+$ survival can be maintained with proton survival below 3\% up to approximately \SI{1.84}{GeV/\clight}. The quoted boundary is obtained by linear interpolation between adjacent simulated momentum points. The localized structure near the proton Cherenkov threshold is sensitive to the tails of the simulated NPE distributions and should not be interpreted quantitatively without dedicated beam data.

As a local robustness check, the PID calculation was repeated with the nearby Rank~2 and Rank~3 optical-calibration solutions in addition to the nominal Rank~1 model. The resulting momentum dependence and the location of the main PID turnover remain qualitatively unchanged. This comparison is used only to verify local stability against nearby calibration solutions and is not interpreted as a quantitative model-uncertainty estimate.

\section{Discussion}
\label{sec:discussion}

\subsection{Measured detector performance}

The beam test directly compares the three radiator configurations. The polished UVT radiator containing bis-MSB at \SI{10}{ppm} by mass gives the largest two-ended photoelectron yield among the tested prototypes. Its SUM response has a measurable longitudinal dependence, but at the nominal 1-photoelectron software-SUM threshold the measured detection efficiency remains above \SI{99}{\percent} throughout the scanned longitudinal range, with no significant degradation over average extracted-particle rates during beam delivery from approximately $2.5\times10^4$ to $2.5\times10^5~\mathrm{s^{-1}}$. These results show that the characterized position dependence is compatible with efficient single-module operation over the tested range.

The comparison of the two no-WLS prototypes isolates the effect of surface treatment at the detector-response level. Roughening the long surfaces increases the collected light relative to polished UVT acrylic without WLS and produces a distinct dependence on hit position and readout side. The measurements therefore show that surface treatment changes both the collected light and the spatial response of the two-ended counter.

\subsection{Optical-model interpretation and limitations}

The data-constrained Geant4 model reproduces the principal response hierarchy of the prototypes and the main longitudinal trends with one common set of optical parameters. Within that description, surface roughening changes the boundary response, whereas bis-MSB introduces wavelength-dependent absorption and re-emission in the bulk acrylic. The comparison is consistent with surface treatment and wavelength shifting altering the collected detector response in qualitatively different ways.

The central-region agreement established in Sec.~5.3 supports the use of the calibrated polished/WLS response for the subsequent PID study, whereas the larger end-region discrepancy limits predictions that depend strongly on the effective rough-surface description. The polished/no-WLS transfer case also shows a larger simulated zero-response component than observed, so predictions that depend on that feature are treated qualitatively or semi-quantitatively.

\subsection{Predictions with fixed model parameters}

After calibration and validation, the optical parameters are held fixed for the detector-design and $K^+/p$ studies. The design comparison shows that changes in acrylic transparency, wavelength shifting, and surface treatment can alter both the absolute light yield and its longitudinal dependence; in particular, the unmeasured rough/WLS configuration is predicted to develop a much stronger position dependence than the adopted polished/WLS configuration. Because this behavior inherits the known limitation of the rough-surface description, its absolute end-to-center variation is not interpreted quantitatively beyond the level supported by the beam-test validation.

The momentum-dependent PID calculation addresses the adopted polished/WLS counter. It is defined for a particle crossing the radiator center at normal incidence and therefore characterizes the intrinsic single-module response at that reference trajectory; it does not represent the full S-2S acceptance. Figure~\ref{fig:pid-momentum-scope} shows that proton suppression is strong at \SI{1.30}{GeV/\clight} for a 99\% kaon-survival working point, while the $K^+$ and proton NPE distributions overlap substantially by \SI{2.00}{GeV/\clight}. For a 90\% $K^+$-survival working point, the nominal model predicts proton survival below 3\% up to approximately \SI{1.84}{GeV/\clight} in the central normal-incidence reference geometry. The location and qualitative behavior of the main turnover are unchanged when the nearby Rank~2 and Rank~3 calibration solutions retained in the present scan are used instead of Rank~1. The localized behavior close to the proton Cherenkov threshold depends on the simulated distribution tails and requires dedicated beam data for quantitative interpretation.

The quantitative calculation in this paper uses $K^+$ and protons. The same optical Cherenkov principle applies to the charge-conjugate $K^-/\bar p$ pair, although a quantitative performance estimate would require a dedicated treatment of species-dependent hadronic interactions and backgrounds.

The polished UVT/WLS configuration is therefore supported by both direct beam performance and the constrained PID extrapolation: it gives the largest measured two-ended light yield among the tested prototypes while maintaining above-99\% detection efficiency across the scanned longitudinal range, and the calibrated model predicts useful $K^+/p$ separation over a finite momentum range beyond the simple binary threshold regime. System-level performance of the installed 16-module detector, including module-to-module response, acceptance weighting, trigger implementation, and experiment-specific backgrounds, will be addressed separately in the S-2S commissioning and physics program.

\section{Summary}
\label{sec:summary}

Three UVT-acrylic Cherenkov-counter prototypes were tested with hadron beams to compare polished acrylic without WLS, roughened acrylic without WLS, and polished acrylic containing bis-MSB at \SI{10}{ppm} by mass. Surface roughening and wavelength shifting produce distinct changes in the measured light yield and longitudinal response. The polished/WLS prototype gives the largest two-ended photoelectron yield among the tested configurations and, at a 1-photoelectron software-SUM threshold, maintains a measured detection efficiency above \SI{99}{\percent} throughout the scanned longitudinal range, with no significant degradation up to an average extracted-particle rate during beam delivery of approximately $2.5\times10^{5}~\mathrm{s^{-1}}$.

A Geant4 optical model constrained by the beam data reproduces the principal ordering and scale of the prototype responses and captures the main longitudinal trends. Over the central $|y|<\SI{200}{mm}$ region, simulated and measured MPV estimators for the calibration configurations typically agree within about 15\%. Larger deviations toward the radiator ends, particularly for rough/no-WLS, mark the limit of extrapolations that depend strongly on the effective rough-surface description.

With the calibrated optical parameters fixed, the model was used to compare unmeasured detector configurations and to quantify the momentum dependence of $K^+/p$ discrimination for the adopted polished/WLS counter. In the central normal-incidence reference geometry, proton survival is predicted to remain below 3\% while retaining 90\% of $K^+$ up to about \SI{1.8}{GeV/\clight}. These results establish the polished UVT/WLS configuration as the preferred design among those studied and provide a beam-constrained quantitative basis for S-2S hadron identification.

\section*{Acknowledgments}
The authors thank the staff of the J-PARC Hadron Experimental Facility for providing the beam time and for their technical support during the beam test. The authors also thank the members of the S-2S Collaboration for valuable discussions and support related to the detector development and beam-test program. This work was supported by JSPS KAKENHI Grant Numbers JP23H00114 and JP26H01412 and was partially supported by SPIRIT2020 of Kyoto University.

\end{document}